\documentclass[%
 reprint,
 amsmath,amssymb,
 aps,
]{revtex4-1}
\usepackage{float}
\usepackage{graphicx}
\usepackage{dcolumn}
\usepackage{bm}
\usepackage{color}
\usepackage{braket}
\usepackage{amsmath,amssymb}
\usepackage{subfig}
\usepackage{textcomp}
\begin{document}

\preprint{APS/123-QED}

\title{Optical Characterization of Disordered Yb-doped Silica Glass Anderson Localizing Optical Fiber}

\author{Cody Bassett$^{1}$}
\author{Matthew Tuggle$^2$}%
\author{Johan Ballato$^{2}$}
\author{Arash Mafi$^{1}$}
\email{mafi@unm.edu}
\affiliation{$^1$Center for High Technology Materials (CHTM) and the Department of Physics \& Astronomy, University of New Mexico, Albuquerque, New Mexico 97131, USA.\\
$^2$Center for Optical Materials Science and Engineering Technologies (COMSET) and the Department of Materials Science and Engineering, Clemson University, Clemson, South Carolina 29625, USA.
}

\date{\today}

\begin{abstract}
We investigate and report the optical and laser characteristics of a ytterbium-doped transverse Anderson localizing optical fiber to develop a fundamental understanding of the light propagation, generation, and amplification processes in this novel fiber. Ultimately, the goal based on the measurements and calculations conducted herein is to design and build a random fiber laser with a highly directional beam. The measurements are based on certain observations of the laser pump propagation and amplified spontaneous emission generation in this fiber. Judicious approximations are used in the propagation equations to obtain the relevant desired parameters in simple theoretical fits to the experimental observations, without resorting to speculations based on the intended construction from the fiber preform.
\end{abstract}

\maketitle

\section{Introduction}
\label{sec:Introduction}

Anderson localization is the absence of wave diffusion in random scattering media~\cite{Anderson1,Anderson2,Anderson1980,wiersma1997localization,transverse-DeRaedt}. In fiber optics, Anderson localization can be observed in the transverse direction to the propagation of light in a fiber with a highly random refractive index profile in the transverse direction but invariant along the fiber~\cite{Mafi-AOP-2015}. The phenomenon observed in an optical fiber is more specifically referred to as transverse Anderson localization (TAL), and the fibers that support it are called Transverse Anderson Localizing Optical Fibers, or TALOFs. The first demonstration of a TALOF was in 2012 by stacking and drawing a mixed bundle of two types of polymer fiber strands~\cite{Mafi-Salman-OL-2012,Mafi-Salman-JOVE-2013}. This fiber was used for detailed studies on the characteristics of TAL in disordered optical fibers, spatial beam multiplexing, and image transport~\cite{Mafi-Salman-OPEX-2012,Mafi-Salman-Multiple-Beam-2013,Mafi-Salman-Nature-2014}. Shortly after, the first silica TALOF was reported by drawing a porous satin quartz glass into a fiber with a transversely random index profile~\cite{Mafi-Salman-OMEX-2012}. Since then, several other groups have also successfully observed TAL in various optical fiber platforms~\cite{chen2014observing,ZHAO:17,Tuan:18}.

A phenomenon closely related to Anderson localization is random lasing. A random laser leverages a highly scattering medium to facilitate feedback in place of a typical cavity~\cite{Cao:05-dev-feat-appl,Wiersma2008-phys-and-appl-random-lasers}. Some of the first observations of random lasing were in a laser dye solution suspended with microparticles~\cite{Lawandy1994,Wiersma1995}, and there have been other reports of random lasing in other disordered media due to diffusive extended modes or Anderson localized modes~\cite{Wiersma1996,Wiersma2008-phys-and-appl-random-lasers}. Ideas behinds TALOFs and random lasers were later converged in a demonstration of random lasing in a dye-filled TALOF--the previously mentioned silica glass-air TALOF was filled with rhodamine-6G dye, where interesting and unique spectral properties were observed~\cite{Mafi-Behnam-Random-Laser-2017}. The laser signal was shown to be highly directional within the Anderson localization regime, with high spectral stability due to the strong mode confinement from localization. The isolated localized modes were also shown to lend to the tunability of the laser.

Dye-filled fibers have great versatility, especially because they can have a broad emission spectrum (typically 30-60\,nm), and different dyes cover a broad wavelength range from the ultraviolet to the near infrared spectrum. However, there are many challenges that one faces in using a dye-filled optical fiber; it is not easy to confine the dye inside the fiber, standard manipulations of the fiber including cleaving and fusion splicing are quite challenging, and it cannot be used for high-power applications where a lot of heat is generated. Moreover, the dye bleaches within minutes of pumping so a dye circulation mechanism must be included~\cite{Gerosa:15}. None of these issues exist for conventional rare-earth-doped silica fibers. Therefore, the next natural step after the proof-of-concept demonstration of lasing in a dye-filled disordered fiber would be to develop a rare-earth-doped version of the fiber, which is reported here, a first to the best of our knowledge. 

\begin{figure}[htp]
    \centering
    \includegraphics[width=8.5cm]{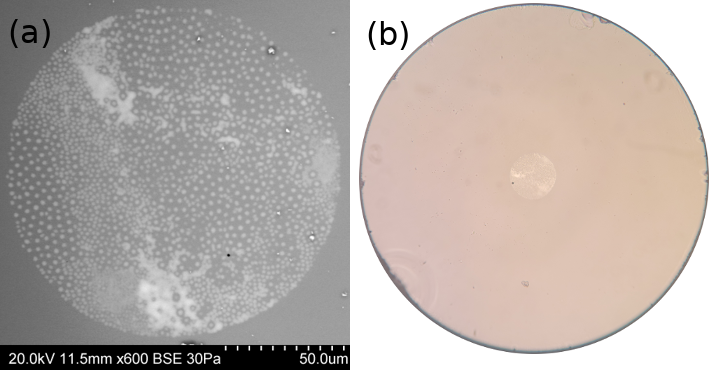}
    \caption{(a) Scanning electron microscope image of the cross-section of the core of the Yb:TALOF. (b) Optical microscope image of the cross-section of the Yb:TALOF, where the core area is visible in the center.}
    \label{fig:Fiber}
\end{figure}
The Yb:TALOF we use in these experiments was drawn using a stack-and-draw method~\cite{Tuggle:20}. This method consists of first drawing out precursor fibers, derived from a mixed powder composition comprising of 58.46\% $\rm{Al_2O_3}$ – 36.54\% SrO – 5.00\% $\rm{Yb_2O_3}$ in mol\%. The precursor fibers are etched down to various diameters with hydrofluoric acid, are bundled into a 5.4\,mm by 6.8\,mm silica tube, are inserted in a 7.8\,mm inner/30\,mm outer diameter preform, and are finally loaded into the draw tower. The stack was then drawn at 1950\,$^{\circ}$C under a vacuum of 15'' Hg to consolidate the core. The preform is drawn to a diameter of roughly 900\,\textmu m, where the resulting core diameter is approximately 140\,\textmu m. During the draw processes, the microcores (the high index phases of glass in the core) are diluted by the surrounding silica, which results in a greater than 25\% decrease in local dopant concentration~\cite{cavillon2019insights}. The resulting index of refraction difference between the high and low index phases is $\Delta n=0.01$, where the high index of refraction is determined to be $n_h=1.4548$ and the low to be $n_l=1.4444$ at 1550\,nm wavelength. This can be regarded as a unification of the molten core draw method and TALOFs.

The final product can be seen in Fig.~\ref{fig:Fiber}, where the bright areas of the core in Fig.~\ref{fig:Fiber}(a) are the high index phases and the darker areas are the low index. The low index component of the core, comprised of 100\% $\rm{SiO_2}$, is the same material used in the cladding. The consistency of the draw is inspected by cleaving a 1\,m sample every 10\,cm and inspecting the cross-sections under a scanning electron microscope. There are slight variations to the microstructure along the fiber, but overall the longitudinal profile of the fiber remains consistent. We have reported previously in Ref.~\cite{Tuggle:20} that localization in Yb:TALOF is easily accomplished in lengths of up to 1\,m using laser sources at 633\,nm and 532\,nm.

A key observation on the Yb:TALOF is that lasing does not happen with the 4\% reflectively from the facets, even when pumping 10-30\,cm long pieces with over 35\,W of power at 975\,nm wavelength. With this observation in mind, we embarked on the investigation presented in this paper to characterize the optical properties of this fiber and understand if it would be amenable to lasing with the right cavity design. There are many questions such as the role and participation of the localized modes versus extended modes in the pump depletion and signal generation processes. While it is conceivable that some of these questions can be answered by extensive modeling and simulation of the disordered fiber, we have taken a much more empirical approach to find answers to these questions. 

By using conventional pump and signal propagation equations in rare-earth-doped fibers and making judicious approximations, we build predictive models for a series of pump-signal measurements based on a minimal set of fitting parameters. The fitting parameters include the resonant and parasitic absorption coefficient of the pump, the pump coupling efficiency into the fiber, the saturation power of the pump, and the number of modes participating in the relevant emission and absorption processes. These parameters allow us to determine the absorption and gain properties of the pump and signal. The outcomes are consistent with a reasonable and physically admissible narrative. The obtained results set the stage for a proper laser design, which will be the subject of a future work. Moreover, the effective parameters obtained from this empirical approach can be compared with a future ab initio modeling and simulation effort, which will likely require extensive computational resources.

We begin this paper by presenting the relevant propagation equations for the pump and the amplified spontaneous emission (ASE) along the fiber. We then present a description of the experimental procedures used for measuring the background absorption of this fiber, with detailed explanation of the specific steps used in order to address the unique challenges of working with this fiber, including its unconventional geometry. Following the background absorption measurements, we describe the procedure used for measuring the pump and ASE output powers. In each stage, we compare the predictions of the theoretical models with the experiments using a minimal set of fitting parameters. Finally, we conclude by examining the consistency of the results and lay out a road map for future work. The Appendix discusses some theoretical details that are used in the paper.
\section{Pump and ASE Propagation Equations}
In this section, we derive the appropriate equations that support our experiments. Because the fiber is novel and a strong element of randomness and disorder is embedded in its construction, it is best to use direct experimentation to obtain as much information about its optical properties as possible without resorting to speculations based on the intended construction from the preform. Our experiments involve mainly pumping various lengths of Yb:TALOF at 975\,nm and measuring the residual output pump power and the generated ASE power. We observe, experimentally, that the output pump power is nearly a linear function of the input power with a slight quadratic correction due to saturation. We also observe a nearly linear behavior for the generated ASE power as a function of the input pump power with a slight quadratic correction. Based on these observations, we derive simplified equations that allow us to extract the properties of the Yb:TALOF.

The rate equation for the upper manifold population density $N_2$ in a quasi-two level system is given by~\cite{becker1999erbium}
\begin{align}
\nonumber
\dfrac{dN_2}{dt}
&=\dfrac{I_p}{h\nu_p}\Big[\sigma^a_p\,N_T-(\sigma^a_p+\sigma^e_p)\,N_2\Big]\\
&+\sum\limits^n_{j=1}\dfrac{I^+_j+I^-_j}{h\nu_j}\Big[\sigma^a_jN_T-(\sigma^a_j+\sigma^e_j)N_2\Big]
-\dfrac{N_2}{\tau_{\rm sp}}.
\label{eq:dN2dt}
\end{align}
The pump frequency (wavelength) is $\nu_p$ ($\lambda_p$). The signal spectrum is sliced into $n$ adjacent segments, where $\delta\lambda$ is the bandwidth for each segment. The signal frequencies and wavelengths are $\nu_j$ ($\lambda_j$), where $j=1,\cdots,n$. The emission and absorption cross sections are defined as
\begin{align}
\sigma^e_j=\sigma^e(\lambda_j),
\quad
\sigma^a_j=\sigma^a(\lambda_j),
\quad
j=1,\cdots,n, \ {\rm or}\ j=p.
\end{align}
The pump and signal local intensities are defined as $I^\pm_j$ for $j=p$ and $j=1,\cdots,n$, respectively. The
$\pm$ superscripts indicate forward and backward propagation, where we consider only the forward pump propagation, because in our experiments the backward propagating pump comes from the reflection at the output facet, which is quite small and can be ignored. Here, $\tau_{\rm sp}$ is the upper manifold lifetime.

In the steady state, where $dN_2/dt=0$, we can solve Eq.~(\ref{eq:dN2dt}) for $N_2$ to obtain
\begin{align}
\label{eq:n2nt}
\dfrac{N_2}{N_T}=
\dfrac{(P_p/P^{\rm sat}_p)\beta_p+\sum\limits_{j,\pm}(P^\pm_j/P^{\rm sat}_j)\beta_j}{1+P_p/P^{\rm sat}_p+\sum\limits_{j,\pm}P^\pm_j/P^{\rm sat}_j},
\end{align}
where we have defined the pump and signal saturation powers as
\begin{align}
\label{eq:saturation}
\dfrac{1}{P^{\rm sat}_p}=\dfrac{\lambda_p\tau_{\rm sp}\sigma^a_p}{hc\beta_pA_p},\qquad
\dfrac{1}{P^{\rm sat}_j}=\dfrac{\lambda_j\tau_{\rm sp}\sigma^a_j}{hc\beta_jA_s}.
\end{align}
$A_p$ is the effective area covered by the pump and $A_s$ is the effective area covered by the spontaneous emission signal, $P_p$ is the total pump power and $P_j$ is the segmented signal power at wavelength $\lambda_j$. Throughout this paper, we assume that the pump and signal have a nearly flat intensity profile, which is also consistent with the highly multi-mode nature of our experiments. We have also defined
\begin{align}
\label{eq:beta}
\beta_k=\beta(\lambda_k)=\dfrac{\sigma^a(\lambda_k)}{\sigma^a(\lambda_k)+\sigma^e(\lambda_k)}.
\end{align}

We can write the ASE (signal) propagation equation as~\cite{becker1999erbium}
\begin{align}
\pm\dfrac{dP^\pm_j}{dz}
=\Big[(\sigma^a_j+\sigma^e_j)N_2-\sigma^a_j N_T\Big]P^\pm_j-\alpha^b\,P^\pm_j+\sigma^e_j\,N_2\,\Pi_j,
\label{eq:dpjdzpj}
\end{align}
where 
\begin{align}
\Pi_j=(\mathbb{V}_j^2/2)hc^2\delta\lambda/\lambda^3_j,
\label{eq:Pidef}
\end{align}
and has units of power. $\mathbb{V}_j$ is the V-number of the fiber at wavelength $\lambda_j$, and $\alpha^b$ is the background loss coefficient of the fiber. The pump propagation equation can similarly be written as
\begin{align}
\label{eq:dppdz}
\dfrac{dP_p}{dz}
=\Big[(\sigma^a_p+\sigma^e_p)N_2-\sigma^a_p N_T\Big]P_p-\alpha^b\,P_p.
\end{align}
In our experiments, both the pump and the ASE cover the entire fiber core and possibly some of the cladding area. Some spatial modes in the fiber are localized in the core due to Anderson localization; however, there are many modes in the fiber that are extended into the cladding with a considerable overlap with the core. Therefore, we anticipate that some modes participating in the propagation of the pump and ASE will have some overlap with the cladding area. This issue requires extensive full-wave electromagnetic modeling of the fiber and is beyond the intended scope of this paper and will be explored in a future publication. For the purposes of this work, we expect $A_p$ and $A_s$ to take values between the core area and the cladding area. One must also be careful with the calculation of the V-number, because it is related to the modal content of the fiber and the effective area covered by the modes. 

For a generic cylindrically symmetric optical fiber, one can write:
\begin{align}
\label{eq:Vnumber}
\mathbb{V}=\dfrac{2\pi R_{\rm eff}}{\lambda}\sqrt{n^2_{\rm co}-n^2_{\rm cl}}
\quad\Rightarrow\quad  
\dfrac{\mathbb{V}^2}{2}=\dfrac{2\pi A_{\rm eff}}{\lambda^2}(n^2_{\rm co}-n^2_{\rm cl}),
\end{align}
where $R_{\rm eff}$ is the effective radius, $A_{\rm eff}=\pi R^2_{\rm eff}$, and $\lambda$ is the wavelength in consideration (pump or ASE). $n_{\rm co}$ and $n_{\rm cl}$ are the higher and lower refractive indices attributed to the core and cladding of the fiber, respectively. $\mathbb{V}^2/2$ represents the total number of modes supported in the effective core. The fiber that we are working with is anything but generic; however, we should still be able to use most of these concepts with the understanding that the calculated values are for the effective parameters that can be compared with generic fibers. The question of whether the light propagation is mainly dominated by Anderson-localized modes or not depends on the measurement of the effective V-number in the absorption and gain processes. If the participating modes are core-localized, they should see an effective refractive index between $n_l$ and $n_h$. Because the high-index doped region of the core is roughly 30\% of the total area of the core, one can roughly estimate that $n_{\rm co}\approx 0.3n_h+0.7n_l$, $n_{\rm cl}=n_l$, and $R_{\rm eff}\approx 70$\,\textmu m. This results in $\mathbb{V}\approx 40$ at 1035\,nm wavelength. Therefore, a V-number on the order of around 40 or a bit larger indicates dominant core-localized mode participation. On the other hand, measuring much larger values for the V-number indicates a substantial cladding mode participation. To appreciate the difference, if one uses the the refractive index of the air surrounding the outer cladding of fiber as $n_{\rm cl}$, and hypothetically uses $R_{\rm eff}\approx 70$\,\textmu m, the resulting V-number is $\mathbb{V}\approx 445$. Because not all cladding modes participate in the absorption and gain process, $R_{\rm eff}$ would certainly come out to be smaller than the outer cladding radius of around 450\,\textmu m, if the mode count is used to determine $R_{\rm eff}$. Therefore, if cladding modes play a strong role, $R_{\rm eff}$ must be chosen larger than the core radius (and smaller than the outer cladding radius), and the V-number must be measured to be larger than 445 (and smaller than 2860). Indeed, our observations confirm the latter scenario and indicate a strong cladding-mode presence in the pump absorption and signal generation and amplification processes. 

Another note of caution here is that the value of $N_T$ should be an area-averaged value in the core because not every point across the fiber core is doped. These points will become more clear as we present the experimental measurements and determine these effective parameters for our fiber.

We first begin by applying Eq.~(\ref{eq:n2nt}) to Eq.~(\ref{eq:dppdz}) to obtain
\begin{align}
\label{eq:PumpEq}
\dfrac{dP_p}{dz} 
=-\alpha^r_p \left[
\dfrac{1+\sum\limits_{j,\pm}(P^\pm_j/P^{\rm sat}_j)(\beta_{j}/\beta_{pj})}{1+P_p/P^{\rm sat}_p+\sum\limits_{j,\pm}P^\pm_j/P^{\rm sat}_j}\right]P_p-\alpha^b\,P_p,
\end{align}
where $\alpha^r_p=\sigma^a_p N_T$ is the resonant absorption coefficient of the pump, and
\begin{align}
\beta_{pj}=\dfrac{\beta_p\beta_j}{\beta_p-\beta_j}.
\end{align}
In Eq.~(\ref{eq:PumpEq}), assuming the pump intensity is much larger than the ASE intensity at any location along the fiber, which is the case in our experiments, we can ignore the ASE contribution and rewrite it as
\begin{align}
\label{eq:alphapr}
\dfrac{dP_p}{dz} 
=-\dfrac{\alpha^r_p}{1+P_p/P^{\rm sat}_p}P_p-\alpha^b\,P_p.
\end{align}
This differential equation can be solved as 
\begin{align}
\label{eq:pumpz}
\dfrac{P_p}{P_{p0}}\left(\dfrac{\alpha_p P^{\rm sat}_p+\alpha^b P_p}{\alpha_p P^{\rm sat}_p+\alpha^b P_{p0}}\right)^{\alpha^r_p/\alpha^b}=e^{-\alpha_p z}.
\end{align}
Here, $P_{p0}$ is the input pump power and $\alpha_p=\alpha^r_p+\alpha^b$. We can approximate the solution to the second order in $P_{p0}$ (technically, second order in $P_{p0}/P^{\rm sat}_p$) and simplify this solution as 
\begin{align}
\label{eq:etaplugedin}
P_p=e^{-\alpha_p z}P_{p0}+\left(e^{-\alpha_p z}-e^{-2\alpha_p z}\right)\dfrac{\alpha^r_p}{\alpha_p} \dfrac{1}{P^{\rm sat}_p}P^2_{p0}
+O(P^3_{p0}).
\end{align}
The first term is the familiar exponential decrease of the pump power due to absorption, while the second term, which is quadratic in $P_{p0}$ provides the first order correction to this formula due to pump saturation. Higher order terms would provide a more accurate account of saturation. It turns out that this second order saturation term is all we need to make a comparison with our experiments. Equation~(\ref{eq:etaplugedin}) can be compared with the experimental measurement of $P_p$ versus $P_{p0}$ to determine $\alpha_p$ (hence $\alpha^r_p$), as well as $P^{\rm sat}_p$.

In a similar manner to Eq.~(\ref{eq:PumpEq}), we can write the ASE propagation equation, Eq.~(\ref{eq:dpjdzpj}), and insert $N_2/N_T$ to write the ASE propagation equation explicitly. We make several reasonable assumptions in order to simplify the calculation; we assume that the ASE signal is quite weak so that it does not saturate. We also use only one spectral slice instead of the ASE summation over the many spectral slices presented earlier, and use $\delta\lambda$ as the effective bandwidth of ASE. Since we make observations of ASE at the output end of the fiber and ASE saturation is ignored, we can study only the forward-propagating ASE and reduce the ASE propagation equation to 
\begin{align}
\nonumber
\dfrac{dP_s}{dz}=&-\alpha^b\,P_s+\alpha^r_s \left[\dfrac{(P_p/P^{\rm sat}_p)(\beta_p/\beta_{ps})-1}{1+P_p/P^{\rm sat}_p}\right]P_s\\
\label{dpsdzSolve}&+g_s\,\dfrac{(P_p/P^{\rm sat}_p)\beta_p}{1+P_p/P^{\rm sat}_p}\,\Pi_s.
\end{align}
The first term on the right-hand side of this equation is the parasitic attenuation, the second term contains both the resonant absorption and amplification, and the last term signifies the spontaneous emission, which seeds ASE. Note that we now use a single subscript $s$ for quantities related to the ASE signal. We also have defined $\alpha^r_s=\sigma^a_s N_T$ and $g_s=\sigma^e_s N_T$.

A comparison of the solutions of Eq.~(\ref{dpsdzSolve}) with measurements of $P_s$ versus $P_{p0}$ allows us to find the value of $\Pi_s$, which is equivalent to finding the effective V-number of the fiber (see Eq.~(\ref{eq:Pidef})), and the number of modes that are participating in the ASE process. Because of the unconventional nature of our disordered fiber, such measurements are essential in determining the effective parameters of the fiber, which would not have been readily calculable without an extensive numerical simulation of the fiber. Finally, we would like to point out that because Eq.~(\ref{dpsdzSolve}) will be used in a single parameter fitting to the measurements, we have decided to use the numerical solution of Eq.~(\ref{dpsdzSolve}) instead of an analytical approximation to find $\Pi_s$.
\section{Experimental Procedure and Data Analysis}
The experimental portion of this paper consists primarily of two different groups of measurements. One is the measurement of the input and output power when illuminating the fiber with a 633\,nm HeNe laser to determine the fiber's parasitic attenuation. The other is pumping the fiber with a laser at 975\,nm wavelength and measuring both the output residual laser power at 975\,nm, or the generated ASE power separately in this part. The latter experiment allows us to determine the resonant absorption coefficient, saturation power, and the number of modes (or V-number) participating in the ASE generation process.

The Yb:TALOF itself is a very thick and rigid fiber, which introduces its own challenges when performing a cutback measurement. In order to remain as consistent as possible, all measurements are done on the same piece of optical fiber. The starting length of the fiber used was 17.5\,cm. When the fiber is cut it must be taken out of the fiber holder in the experimental setup, but it is always cut from the output end to maintain the input facet. What follows is a brief overview of the experimental procedure, with greater detail of the procedure in each subsection. First, the fiber is placed in the setup used with the HeNe laser. Measurements are taken of the input and output power for several different values of input powers. When one "sweep" of the input power is completed, the fiber is transversely realigned to slightly alter the coupling into the core of the fiber, and the "sweep" of measurements is taken again to obtain appropriate statistics. 

Once all measurements for the HeNe are completed for a given length of fiber, the fiber is then removed and placed in the setup for the 975\,nm laser. For the second set of measurements, where the fiber is subjected to the 975\,nm laser, the input pump power is measured followed by a measurement of the output pump power, now with a short-pass edge filter in place to remove ASE generation from the measurement. Once the output pump power has been measured, the filter is exchanged with a long-pass edge filter to now only allow ASE generation to pass through and the ASE power is measured. The input pump power is then increased, and the ``sweep'' and realigning procedure continues in a similar manner to the HeNe laser experiment. Once this is completed with both the HeNe laser and the 975\,nm pump, the fiber is removed from the setup and is cleaved 2.5\,cm from the output end of the fiber. The now 15\,cm length of fiber is placed back in the HeNe setup, and the entire process is repeated again. This was done for six different lengths of fiber, from 17.5\,cm down to 5\,cm, in 2.5\,cm increments. 

\subsection{Measurement of Background Absorption}
\label{sec:bg}

\begin{figure}[htp]
    \centering
    \includegraphics[width=8.5cm]{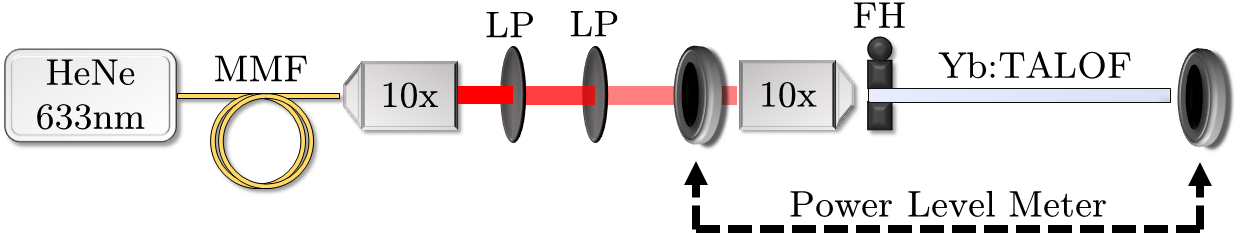}
    \caption{HeNe Experimental diagram, where MMF is a multi-mode fiber (50\,\textmu m diameter, 0.22 NA), LP is a linear polarizer, and FH is the fiber holder.}
    \label{fig:HeNeDiagram}
\end{figure}

We begin the characterization of this fiber with a measurement of the background absorption. As mentioned in Section~\ref{sec:Introduction}, this fiber presents a challenge in performing a cutback measurement, as it is a solid silica fiber with a nearly 1\,mm diameter cladding. It is not possible to leave the fiber in a fiber holder and cleave the output end at the same time, so we must alter the experimental setup in such a way that we can remove and replace the fiber for cleaving with minimal change to the position of the fiber along the optical axis. This is done by placing the input facet of the fiber flush with the front edge of the fiber holder (see Fig.~\ref{fig:HeNeDiagram}). This is easy to accomplish, as the fiber has a large diameter and can be easily seen with a small magnifying lens. The result is that the fiber remains at nearly the exact same distance from the coupling objective when the fiber is removed and replaced after cleaving. 

To measure background loss, we illuminate the fiber with a laser wavelength far outside the absorption bandwidth of ytterbium (which tapers off near 850\,nm). We use a 633\,nm HeNe laser with a single-mode output of 10\,mW, couple the light into a multi-mode optical fiber, and re-collimate the output in order to obtain a highly multi-mode beam profile to fill the entire fiber core. This is done for two reasons; one is that having a higher mode profile will increase the coupling efficiency to the highly multi-mode core of the Yb:TALOF, and the other being that our 975\,nm pump diode is fiber coupled to a multi-mode fiber and thus also has a higher-mode beam profile. The 975\,nm pump source fiber has a 105\,\textmu m diameter and 0.22 NA, so in order to best match the V-number for both laser beam profiles, we couple the HeNe into a 50\,\textmu m diameter and 0.22 NA fiber. The resulting V-number of the fiber delivering the HeNe laser input is 54.6, while the V-number of the fiber delivering the 975\,nm pump is 74.4. 

To prepare the initial alignment of the setup, we first place the fiber flush with the fiber holder and couple into the core of the fiber. This is checked by imaging the output facet of the fiber with an objective onto a small screen nearly 30\,cm away and ensuring that the laser light is not strongly coupled to the cladding. Then, we adjust the optical axis so that the beam spot size is roughly the same size as that of the core. This can be verified when adjusting the transverse alignment of the fiber, and observing that little movements in any direction begins to couple light into the cladding of the Yb:TALOF. The initial transverse alignment of the fiber (or its rotation about the optical axis when replaced after cleaving) is not critical, so long as the pump light is completely coupled into the core because the transverse profile of the fiber is purposely realigned between steps of the experiment later on. It is, however, important to ensure that the fiber is being pumped from the same side every single measurement, as we have observed that pumping the Yb:TALOF from one side can yield slightly different results than pumping the opposite side. A simple ink mark on the side of the fiber near the input end was used for distinction.

The experimental procedure primarily consists of aligning the transverse profile of the fiber so that it couples directly into the core, then measuring the input pump power (before the objective that couples into the fiber) and the output pump power directly at the end of the fiber. For each fiber length, a sweep of five different power levels, each ranging from 100\,\textmu W to 8\,mW, is conducted five times for a total number of twenty-five measurements. The HeNe power is attenuated through the use of two linear polarizers. During each sweep of the input power, we ensure that the fiber is not moved in any way. Once the first set of five measurements is complete, the fiber is then transversely misaligned and realigned to couple back into the core. This is done so that the coupling for each set of measurements is always slightly different from one another, and provides a meaningful average for the analysis of the fiber. After realignment, the sweep is done again, and this is repeated for a total of five sweeps. Once the total group of twenty-five measurements is completed, the fiber is then removed and placed in the experimental setup for the 975\,nm pump/ASE measurements. After the pump/ASE measurements are completed, the fiber is cleaved 2.5\,cm shorter, then the group of twenty-five measurements for HeNe is repeated again. This is done for all six lengths of fiber. 

\begin{figure}[htp]
    \centering
    \includegraphics[width=8.5cm]{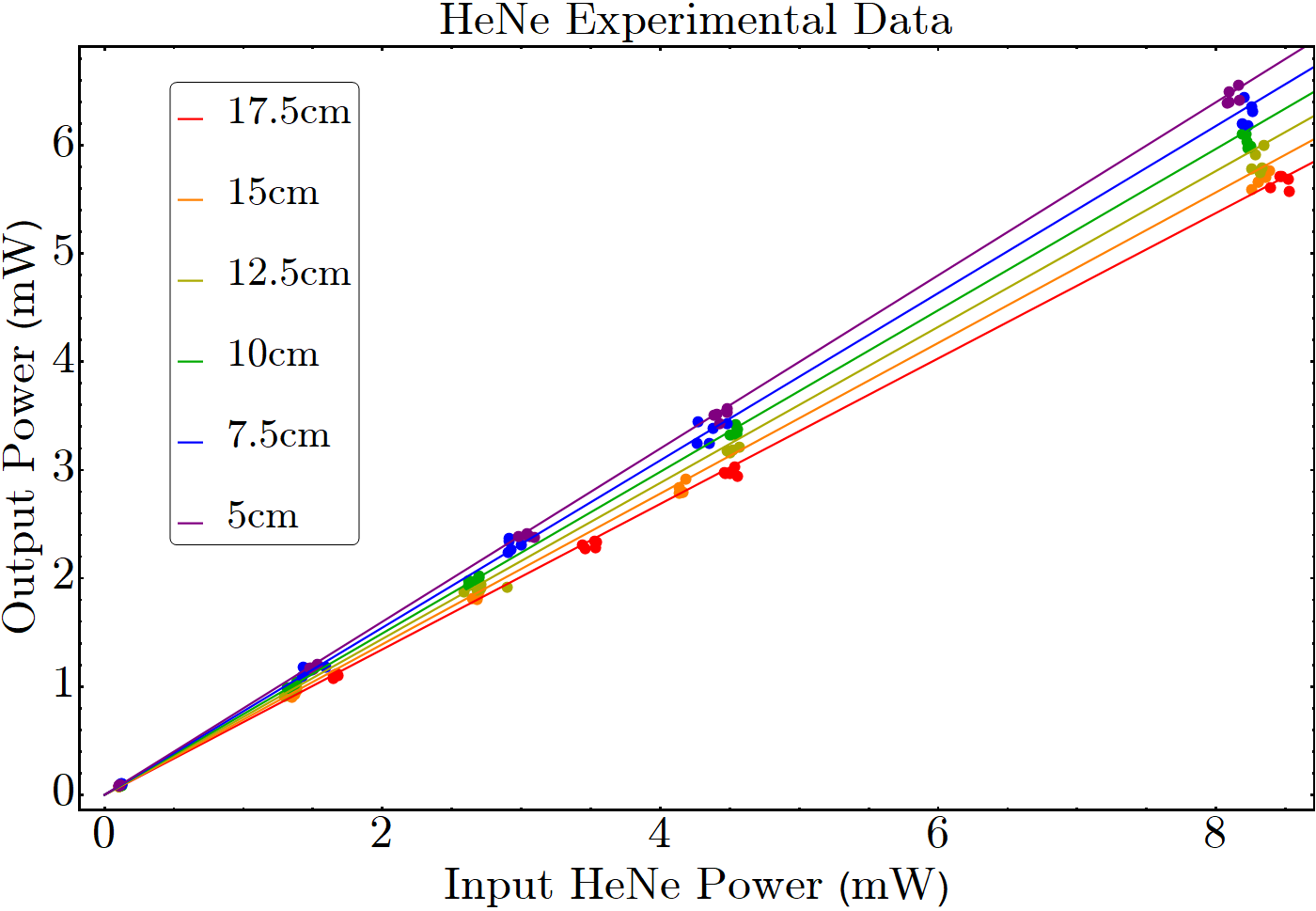}
    \caption{Plotting of the experimental data when measuring the output versus the input power when pumping the Yb:TALOF with 633\,nm HeNe light. Each color represents a different length of fiber. The plot can be seen in color online.}
    \label{fig:HeNeData}
\end{figure}

Plotting the output power as a function of the input power shows a linear trend. In Fig.~\ref{fig:HeNeData} each color represents a different fiber length, and each line represents the best linear fit for the data of each fiber length. 
The output versus input HeNe power must follow the Beer–Lambert law 
\begin{align}
\label{eq:HeNESolution}
P_p=P_{p0}\epsilon_h e^{-\alpha^b z},
\end{align}
where $z$ is length of the fiber, $\alpha^b$ is the background absorption associated with parasitic loss, and $\epsilon_h$ is the coupling efficiency of the HeNe light into the fiber. We use Mathematica to perform a nonlinear model fit to all HeNe data simultaneously using Eq.~(\ref{eq:HeNESolution}). We ensure that all power levels have the same weight in the fitting procedure by using the normalized ratio $P_p/P_{p0}$. The nonlinear model fit allows us to simultaneously solve for the coupling efficiency and the background loss coefficients at the same time. We find that the coupling efficiency for the HeNe is $\epsilon_h=0.858\pm0.006$ and the background absorption of the fiber is $\alpha^b=1.39\pm0.06\,\rm{m}^{-1}$. The fiber tends to exhibit sub-optimal coupling efficiency, which is in agreement with our previous predictions of coupling efficiency for a TALOF~\cite{Mafi-gvd-Anderson-2019}.

\subsection{Solving the Pump Propagation Equation}
\label{pump}

\begin{figure}[htp]
    \centering
    \includegraphics[width=8.5cm]{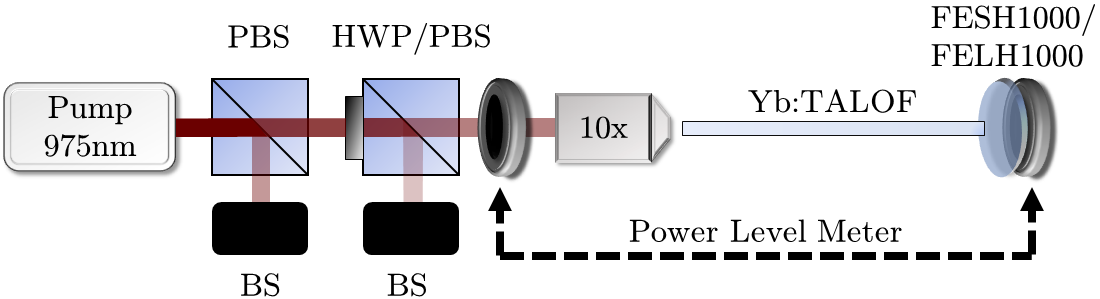}
    \caption{Pump and ASE Experiment Diagram, where PBS is a polarizing beam splitter, HWP is a half-wave plate, BS is a beam stop, and FESH1000/FELH1000 are edge-pass cut-off/cut-on filters, respectively.}
    \label{fig:PumpDiagram}
\end{figure}
Following the measurements in the HeNe setup, we place the fiber in the 975\,nm experimental setup for measurements of the pump and ASE. Like before, it is important to ensure that the fiber is placed flush with the fiber holder for all measurements and that the fiber is being coupled into the same facet each time the fiber is replaced after being cleaved. Alignment follows the same procedure as with the HeNe setup, and measurements are also taken in a similar manner. A sweep of five pump powers ranging from 100\,mW to 9\,W is done five times for each length of fiber. In this setup, because we are using a high-power laser, we attenuate the beam using a combination of a polarizing beam splitter followed by a half-wave plate and another polarizing beam splitter (See Fig.~\ref{fig:PumpDiagram}). First we measure the input and output pump power at the pump wavelength. When measuring the output, we attach a FESH1000 short-pass filter (fabricated by Thorlabs) to the power meter to filter out all light with a wavelength longer than 1000\,nm. Following the measurement of the pump output, we replace the short-pass filter with a FELH1000 long-pass filter to remove all light with a wavelength less than 1000\,nm and measure the ASE output power peaked at around 1035\,nm. As mentioned before, between each sweep, there is a transverse realignment to the core of the fiber.

\begin{figure}[htp]
    \centering
    \includegraphics[width=8.5cm]{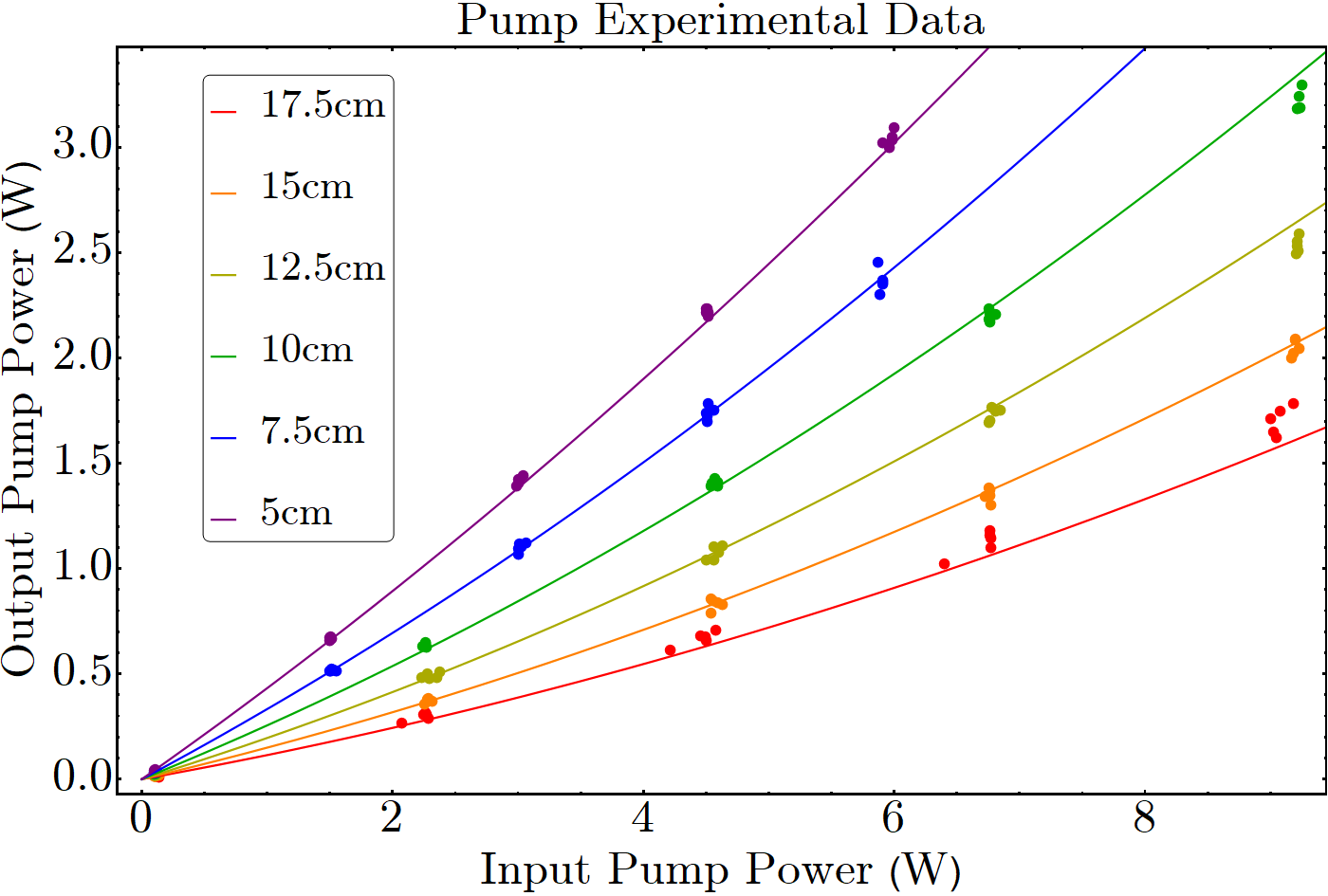}
    \caption{Plotting of the experimental data when measuring the output versus the input power when pumping the TALOF with 975\,nm pump light. Each color represents a different length of fiber.}
    \label{fig:PumpData}
\end{figure}

We see in Fig.~\ref{fig:PumpData} a clear quadratic relationship, which was the main motivating factor in deriving Eq.~(\ref{eq:etaplugedin}).
We perform a nonlinear model fit to a quadratic function with all of the data, again weighting the function by the input pump power. We must include a coupling efficiency, $\epsilon_p$, for the input pump power, so the equation used for nonlinear fitting is  
\begin{align}
\label{eq:Pumpsolution}
P_p=\epsilon_p P_{p0} e^{-\alpha_p z}+\left(e^{-\alpha_p z}-e^{-2\alpha_p z}\right)\dfrac{\alpha^r_p}{\alpha_p} \dfrac{1}{P^{\rm sat}_p}\epsilon_p^2 P^2_{p0}.
\end{align} 
The best fit parameters are a coupling efficiency of $0.721\pm0.005$ for the pump at 975\,nm, a resonant absorption of $9.4\pm0.1\,\rm{m}^{-1}$, and a saturation power of $7.7\pm0.2\,\rm{W}$. For $\alpha^b$, we use the value measured in the HeNe experiment. Using the resonant absorption coefficient, we can now also determine the ion density of the core using the relation $\alpha_p^r=N_T \sigma_p^a$, where $N_T$ is the area-averaged ytterbium ion density and $\sigma_p^a$ is the absorption cross section of ytterbium at the pump wavelength. The effective ion density is roughly $(3.62\pm 0.09)\times10^{24}\,\rm{m}^{-3}$. The doped region of the core is roughly 30\% of the total area of the core (see Fig. \ref{fig:Fiber}), so the actual ion density of the doped regions must be $(1.22\pm 0.03)\times10^{25}\,\rm{m}^{-3}$.

We can use the fitted value of the saturation power in Eq.~(\ref{eq:saturation}) to determine the value of the effective pump area, $A_p$. The effective pump radius, $r_p=\sqrt{A_p/\pi}$ is determined to be $234\pm 3$\,\textmu m, which is consistent with our notion that the pump propagates both in core-localized modes confined in the core radius of around 70\,\textmu m and some cladding modes with large core overlap. The corresponding effective V-number can is be calculated to be $V_{\rm eff}=1578\pm 20$. These values will be compared with those obtained from ASE measurements in the next section.
\subsection{Solving the ASE Propagation Equation}
\label{sec:ase}
\begin{figure}[htp]
    \centering
    \includegraphics[width=8.5cm]{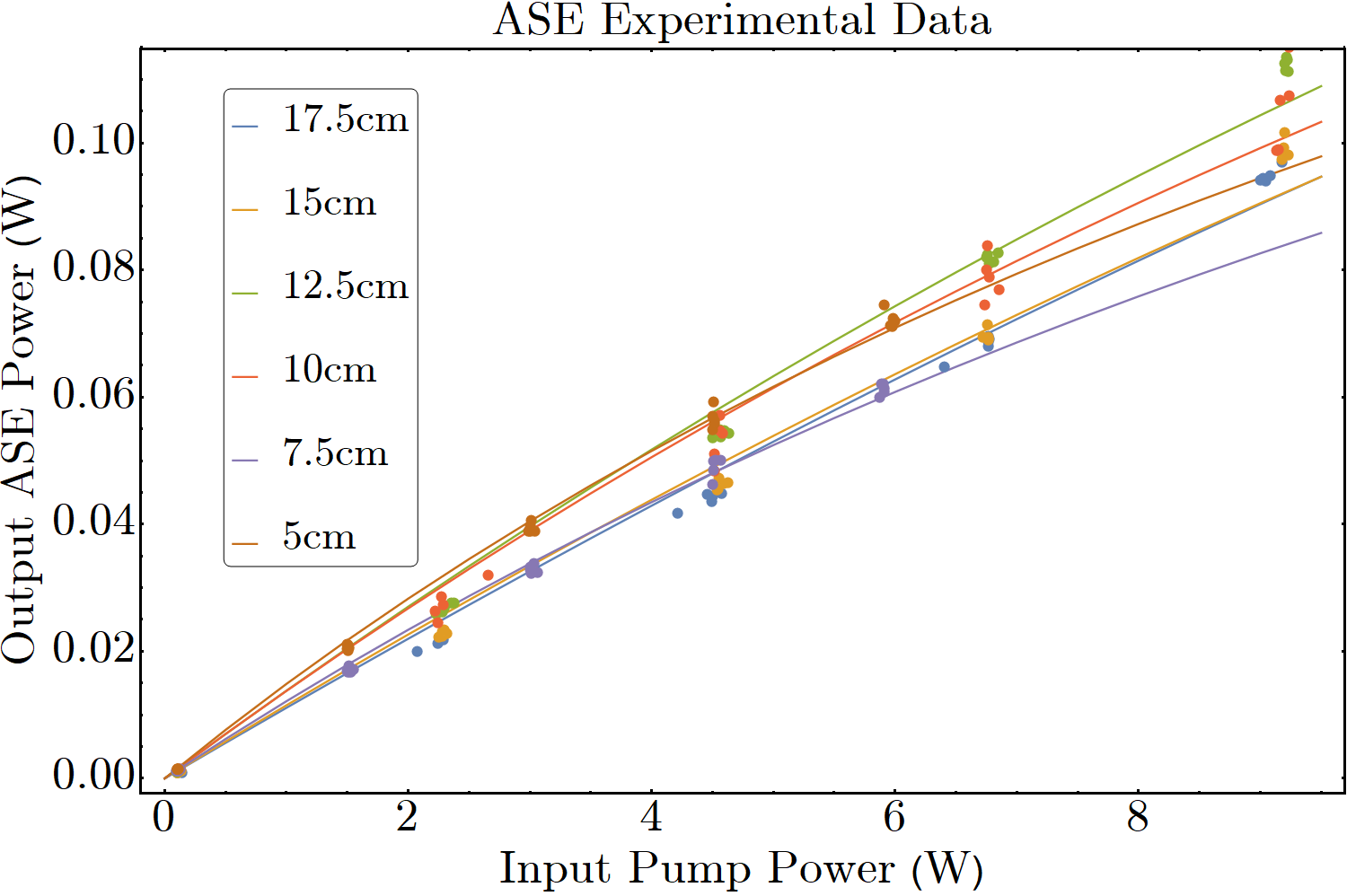}
    \caption{Plotting of the experimental data when measuring the output ASE generation versus the input pump power when pumping the TALOF with 975\,nm pump light. Each color represents a different length of fiber.}
    \label{fig:ASEData}
\end{figure}

The measurement of the output ASE power takes place immediately after the measurement of the 975\,nm pump output power at each step and the details of the procedure are described in Section~\ref{pump}. The output ASE power versus the input pump power for different lengths of the fiber are shown in Fig.~\ref{fig:ASEData}.
The continuous lines represent the best fits by numerically solving the differential equation~(\ref{dpsdzSolve}) with $P_p$ replaced with the solution presented for the pump power in Eq.~(\ref{eq:Pumpsolution}). We note that including the second order term in $P_{p0}/P^{\rm sat}_p$ is important for getting a good fit to the measured data using Eq.~(\ref{dpsdzSolve}). The differential equation is solved for each fiber length individually, using the value of $\Pi_s$ as the only fitting parameter. We observe that as the fiber is cut shorter from 17.5\;cm to 5\;cm, the calculated value of $\Pi_s$ for the best fit increases from 1.6\,W to 4.0\,W, respectively. The average value of $\Pi_s$ is calculated as $2.4\pm0.9$\,W. This translates to a mean value V-number of $1484\pm278$, or $(11\pm4)\times10^5$ guided modes. The V-number is calculated by assuming the ASE bandwidth, $\delta\lambda$, is approximately 40\,nm. Note that the obtained V-number from the ASE measurements is consistent with that obtained from the pump power measurements once the inverse relation between the V-number and wavelength is taken into account. A plausible explanation for the inverse relationship between the value of $\Pi_s$ and the fiber length is that as the fiber length is increased, the strong side scattering loss results in a smaller number of modes being pumped and participating in the ASE generation and amplification process. Having said that, the obtained range for the V-number is consistent with that obtained in  Section~\ref{pump} for the pump. 

As a final note, we would like to point out that the spontaneous emission lifetime of ytterbium in the Yb:TALOF used in the calculations has been measured with the same methods used for a typical optical fiber; a short fiber segment is pumped with a low-power 975\,nm laser light source after passing through a beam chopper, and the side emission from the fiber is measured using a fast-response photodiode. The light collection to the photodiode is low enough not to saturate the detector. Exponential decay is fitted to the photodiode response, giving us a spontaneous emission lifetime of 0.86\,ms. 
\section{Conclusion}
We report the measurements of gain and loss parameters of a Yb-doped TALOF. The fiber is non-standard, so an empirical approach is used by combining direct measurements and semi-analytical approximations to determine the relevant optical and laser properties of fiber. 
We measure the background absorption using a HeNe laser, which is determined to be $\alpha^b=1.39\pm0.06\,\rm{m}^{-1}$. We use an analytical solution for the pump propagation equation that matches the observed results of a quadratic relation between the input and output pump power at 975\,nm, and fit the data to the solution to find the best fit for the pump saturation power and resonant absorption coefficient, $P_p^{\rm{sat}}$ and $\alpha_p^r$, respectively. We determine that the pump saturation power is $7.7\pm0.2\,\rm{W}$ and the pump resonant absorption coefficient is $9.4\pm0.1\,\rm{m}^{-1}$. The effective ion density, derived from the resonant absorption, is roughly $(3.62\pm 0.09)\times10^{24}\,\rm{m}^{-3}$. A numerical solution is used for the ASE propagation equation with a single fitting parameter, $\Pi_s$, which is determined to be $2.4\pm0.9$\,W. These parameters complete the set required for a laser or amplifier design based on this fiber, which will be the subject of a future publication.

Future work will include designing a laser cavity based on Yb:TALOF and comparing the performance of the laser with that expected from the empirically obtained parameters for the fiber. It will also include lowering the parasitic absorption, $\alpha^b$, to improve the laser characteristics. Ultimately, the goal is to design and build a random fiber laser with a highly directional beam and possibly a broad bandwidth based on Yb:TALOF.
\section{Funding}
A. Mafi and C. Bassett acknowledge support by Award Number 1807857, and J. Ballato and M. Tuggle acknowledge support by Award Number 1808232 from the National Science Foundation (NSF).
\section{Acknowledgments}
M. Tuggle is now with Corning Incorporated.\\[3mm]

\providecommand{\noopsort}[1]{}\providecommand{\singleletter}[1]{#1}%

\appendix
\section*{Appendix: Generation of spontaneous emission}
This Appendix presents the derivation of $\Pi_s$ used in Eq.~\ref{dpsdzSolve} for the generation of spontaneous emission. The material presented in here is not novel and is only intended as a coherent review of the topic for the interested reader. The presentation is largely adapted from Ref.~\cite{Saleh-Teich}.

In a three dimensional cubic space of edge length $d$ with planar-mirror boundaries, The number of modes lying in the frequency interval between $0$ and $\nu$ are given by $8\pi n^3\nu^3d^3/(3\,c^3)$. The density of modes $\mathbf{M}(\nu)$ is defined as the number of modes per unit volume lying in the frequency interval between $\nu$ and $\nu+d\nu$, which is given by 
\begin{align}
\mathbf{M}(\nu)=\dfrac{1}{d^3}\dfrac{d}{d\nu}\left(\dfrac{8\pi n^3\nu^3d^3}{3\,c^3}\right)=\dfrac{8\pi n^3\nu^2}{c^3}.
\end{align}
In a cavity of volume $V$, the probability density (per second) for spontaneous emission into one optical frequency mode $\nu$ is given by
\begin{align}
\label{eq:counts0}
\mathfrak{p}_{\rm sp}=\dfrac{c/n}{V}\sigma^e(\nu),
\end{align}
where $c/n$ is the speed of light in a medium of refractive index $n$, and $\sigma^e(\nu)$ is the value of the emission cross section of ytterbium at frequency $\nu$. The overall spontaneous emission probability density per time is given by multiplying the probability of spontaneous emission into a single mode $\nu$ by the density of modes and the volume of the medium, and then integrating over all frequencies
\begin{align}
\nonumber
\mathbb{P}_{\rm sp}&=\int_0^\infty\left[\dfrac{c/n}{V}\sigma^e(\nu)\right]\left[V\mathbf{M}(\nu)\right]d\nu\\
&=\dfrac{c}{n}\int_0^\infty\sigma^e(\nu)\mathbf{M}(\nu)d\nu.
\end{align}
It must be noted that $\sigma^e(\nu)$ is a sharply peaked function and $\mathbf{M}(\nu)$ does not change appreciably over the region where $\sigma^e(\nu)$ is non-zero. Given this, we can pull $\mathbf{M}(\nu)$ out of the integral as a constant $\mathbf{M}(\nu_0)$, where $\nu_0$ is the peak frequency of $\sigma^e(\nu)$. We then have 
\begin{align}
\mathbb{P}_{\rm sp}=\dfrac{c}{n}\,\mathbf{M}(\nu_0)\int_0^\infty\sigma^e(\nu)d\nu.
\end{align}
The inverse of the spontaneous emission probability, $1/\mathbb{P}_{\rm sp}=\tau_{\rm sp}$, is defined as the spontaneous lifetime. The quantity $\int_0^\infty\sigma^e(\nu)d\nu=S$ is called the transition strength. The lineshape function $g(\nu)$ is defined such that $g(\nu)=\sigma^e(\nu)/S$, where $\int_0^\infty g(\nu)d\nu=1$. We can put all this information together and obtain
\begin{align}
\label{eq:counts1}
\dfrac{1}{\tau_{\rm sp}}&=\left[\dfrac{c}{n}\right]\left[\dfrac{8\pi n^3\nu^2}{c^3}\right]S=\dfrac{8\pi n^2\nu^2}{n^2}S\\
\nonumber&\Longrightarrow \quad S=\dfrac{\lambda^2}{8\pi n^2 \tau_{\rm sp}}.
\end{align}

We can then use $\sigma^e(\nu)=S g(\nu)$ to write
\begin{align}
\label{eq:counts2}
\sigma^e(\nu)=\dfrac{\lambda^2}{8\pi n^2 \tau_{\rm sp}}g(\nu).
\end{align}
We can rewrite Eq.~(\ref{eq:counts2}) as
\begin{align}
\dfrac{1}{\tau_{\rm sp}}g(\nu)=\dfrac{c}{n}\mathbf{M}(\nu)\sigma^e(\nu).
\end{align}
The probability density per second that an atom in the upper level spontaneously emits a photon of frequency between $\nu$ and $\nu+d\nu$ is
\begin{align}
\widetilde{P}_{\rm sp}(\nu)\,d\nu=\dfrac{1}{\tau_{\rm sp}}g(\nu)\,d\nu=\dfrac{c}{n}\mathbf{M}(\nu)\sigma^e(\nu)d\nu.
\end{align}
We can immediately conclude that $\mathbb{P}_{\rm sp}=\int_0^\infty\widetilde{P}_{\rm sp}(\nu)d\nu$.
If we multiply $\widetilde{P}_{\rm sp}(\nu)\,d\nu$, the photon number per second, by the volume density of atoms in the upper level and the energy of a photon, we obtain the generated spontaneous emission power density at the frequencies between $\nu$ and $\nu+d\nu$
\begin{align}
\label{Eq:pdensity}
N_2\,h\nu \dfrac{c}{n}\,\mathbf{M}(\nu)\,\sigma^e(\nu)\,d\nu.
\end{align}

To calculate the fraction of spontaneous emission that is coupled into a multi-mode fiber, we must consider the numerical aperture of the fiber in order to determine what fraction out of the entire $4\pi$ solid angle is coupled into guided modes of the fiber. The solid angle of a cone with apex angle $2\theta$ is given by
\begin{align}
\Omega=2\pi(1-\cos\theta)\approx \pi\theta^2.
\end{align}
In a multi-mode optical fiber, $\theta$ is nearly equal to the numerical aperture of the fiber. Inside the optical fiber, this angle changes according to the Snell's law. This angle changes to $\theta/n$, where $n$ is the nominal refractive index of the optical fiber. The fraction of the spherical radiation that makes it to the right-moving fiber modes is given by $(\pi\theta^2/n^2)/(4\pi)=\theta^2/(4n^2)$. 

Recall that Eq.~(\ref{Eq:pdensity}) applies to all frequency modes in a homogeneous medium of refractive index $n$, with emission into all $4\pi$ steradian. Therefore, Eq.~(\ref{Eq:pdensity}) must be rescaled by the factor of $\theta^2/(4n^2)$ to account only for the radiation that makes it to the right-moving fiber-guided spontaneous emission. To obtain the differential power generated in volume $V=Adz$, we need to multiply also by $V=Adz$. Note that a factor of 2 for polarization multiplicity is already included in $\mathbf{M}(\nu)$. We obtain
\begin{align}
\label{eq:dpspdz}
\nonumber
\dfrac{dP_{\rm sp}}{dz}&\sim N_2\,h\nu \dfrac{c}{n}\,\mathbf{M}(\nu)\,\sigma^e(\nu)\,\delta\nu \dfrac{A \theta^2}{4n^2}\\
&\sim (h\nu\,\delta\nu) N_2 \sigma^e(\nu) \dfrac{2\pi A \theta^2}{\lambda^2},
\end{align}
where we have used $\mathbf{M}(\nu)=n^38\pi/(c\lambda^2)$.

Now, recall that the V-number of a fiber is approximately given by $\mathbb{V}_s\approx (2\pi a/\lambda_s)\theta$, where $a$ is the core radius. We can therefore write $\mathbb{V}_s^2\approx 4\pi A\theta^2/\lambda^2$, where $A$ is the area of the core. We can rewrite~\ref{eq:dpspdz} as
\begin{align}
\label{eq:dpspdz2}
\dfrac{dP_{sp}}{dz}&\sim \left(\dfrac{\mathbb{V}_s^2}{2}h\nu\,d\nu\right) N_2 \sigma^e(\nu)\sim\left(\dfrac{\mathbb{V}_s^2}{2}\dfrac{hc^2\delta\lambda}{\lambda^3}\right) N_2 \sigma^e(\nu),
\end{align}
which explains why $\Pi_s$ is defined as $(\mathbb{V}_s^2/2)hc^2\delta\lambda/\lambda^3_s$. For single mode optical fibers, a similar line of argument can be pursued and the final result is identical to Eq.~(\ref{eq:dpspdz2}), except $\mathbb{V}_s^2/2$ is replaced with 2. This is consistent with the fact that the total number of modes in a multi-mode optical fiber is approximately given by $\mathbb{V}_s^2/2$, while a single mode fiber supports two polarization modes.
\end{document}